\documentclass[10pt]{revtex4}
\usepackage{hyperref}
\usepackage{amsmath}
\usepackage[psamsfonts]{amssymb }
\usepackage{mathrsfs}

\begin{document}

\title{\Large Conformal to Harmonic Gauge for Bosonic Strings}

\author{Vipul Kumar Pandey\footnote {e-mail address: vipulvaranasi@gmail.com}}
\author{ Bhabani Prasad Mandal\footnote {e-mail address: bhabani.mandal@gmail.com}}

\affiliation {Department of Physics, 
Banaras Hindu University, 
Varanasi-221005, INDIA.}

\begin{abstract}
We consider Polyakov theory of Bosonic strings in conformal gauge which are used to study conformal anomaly. However it exhibits ghost number anomaly. We show how this anomaly can be avoided by connecting this theory to that of in background covariant harmonic gauge which is known to be free from conformal and ghost number current anomaly, by using suitably constructed finite field dependent BRST transformation.
\end{abstract}

\maketitle
\section{Introduction}\label{Introduction}
Bosonic string in path integral formulation\cite{1,2,3} has been studied in various gauges. The simplest choice among such gauges is the conformal gauge\cite{1,3,4,5} which has been used to study many important properties of bosonic strings. In conformal gauge, upon renormalization the effective action which is defined as a whole by one loop Feynman diagram, exhibits conformal anomaly\cite{6}. Besides the conformal anomaly, there is another important anomaly associated with conformal gauge is the ghost number current anomaly on curved world-sheet\cite{5}. Bosonic string has also been investigated in detail in a bit complicated harmonic gauge\cite{7}(a choice similar to Lorenz gauge in QED). In harmonic gauge the standard $(D-26)$ factor appears naturally\cite{1}. Even though harmonic gauge is bit difficult to use for quantization it does not require the ghost field interaction at the vertex. Further ghost number current anomaly doesn't appear in the background covariant harmonic gauge\cite{8} but the absence of a ghost number anomaly is achieved at the expense of a new anomaly in the sector involving the Nakanishi-Lautrup field\cite{9}. The BRST analysis of the bosonic string in the this perspective have been discussed in\cite{10}.

BRST quantization \cite{11} is an important and powerful technique to deal with a system with constraints \cite{12}. It enlarges the phase space of a gauge theory and restores the symmetry of the gauge fixed action in the extended phase space keeping the physical contents of the theory unchanged. Recently BRST quantization has been used to study some new properties in various theories like string theories \cite{14,15,20,27,24}, superstring theories \cite{13,16,17,18,19,21,22,23,25,26}, M theory \cite{28,30}, Chern-Simons theory \cite{29,32,34} and ABJM theory \cite{31,33,35}. We indicate how various BRST invariant effective theories are interlinked by considering the finite-field-dependent version of the BRST(FFBRST) transformation, introduced by Joglekar and Mandal \cite{36} about two and half decade ago. FFBRST transformations are the generalization of the usual BRST transformation where the usual infinitesimal, anti-commuting, global transformation parameter is replaced by a field-dependent but global and anti-commuting parameter. Such generalized transformation protects the nilpotency and retains the symmetry of the gauge fixed effective actions. The remarkable property of such transformations are that they relate the generating functionals corresponding to different effective actions. The non-trivial Jacobian of the path integral measure under such a finite transformation is responsible for all the new results. In virtue of this remarkable property, FFBRST transformations have been investigated extensively and have found many applications in various gauge field theoretic systems \cite{37,38,39,40,41,42,43,44,45,46,47,48,49,50}. A similar generalization of the BRST transformation with the same motivation and goal has also been carried out more recently in a slightly different manner \cite{51} where a Jacobian contribution for such transformation is calculated without using any ansatz. Recently FFBRST transformation has been used successfully in some models of string theory\cite{52,53,54,55}.

In the present work we consider Polyakov action in path integral formulation in the conformal gauge where ghost number current anomaly in curved world-sheet is present. Constructing appropriate finite field dependent BRST transformation we connect the generating functional in background covariant(bc), conformal to harmonic gauge in bosonic string. In this way we will avoid the ghost number current anomaly. 

Now we present the plan of the paper. In the next section we discuss bosonic string in different gauges and write BRST invariant effective action. Connection between the generating functional in bc harmonic gauge and conformal gauge presented in sec. III. The last section is kept for conclusion. 
  
\section{Bosonic String Action}\label{Bosonic String Action}

In the path integral formalism of bosonic string the Polyakov action is written as\cite{1}
\begin{equation}
S_0 = \int d^2x\frac{1}{2}\sqrt{-g} g^{ab}\partial_a X^\mu \partial_b X_\mu 
\label{bsa}
\end{equation}
where $X^\mu, \mu = 0,1,2....D-1$ is string co-ordinate and $g^{ab}, a,b = 0,1$ is world-sheet metric.
This action is invariant under both diffeomorphisms and Weyl transformations in following ways
\begin{equation}
g_{ab}\rightarrow g'_{ab} = g_{ab} + \nabla_a \xi_b + \nabla_b \xi_a, \quad g_{ab}\rightarrow g'_{ab} = (1 + 2\sigma) g_{ab}
\label{gab}
\end{equation}
To factor out the infinite factors in the functional integral associated with these transformations usually we choose the gauge condition and introduce the Jacobian for the change of the gauge condition under the infinitesimal diffeomorphisms ($\xi^a$) and Weyl transformations ($\sigma$). There are various gauge choices for this theory. In the present work we have restricted ourself to two important gauges namely conformal gauge and background covariant harmonic gauge. The BRST invariant effective actions in these gauges for Polyakov action are discussed below. 

\subsection{Conformal Gauge}\label{Conformal Gauge}

Conformal gauge has been used extensively in the discussion of various problems. This gauge is very useful in introducing Weyl symmetry and in renormalizing the theory \cite{1,3,4,5}.The conformal gauge condition is expressed as $ h_{ab} (\equiv g_{ab} -  \hat{g}_{ab} = 0)$ and is incorporated into the following gauge-fixing and FP ghost term in a BRST invariant manner\cite{4},

\begin{eqnarray}
\mathscr{L}_{cf} = \lambda (\mathscr {L}_{gf} + \mathscr {L}_{gh}) = -i \delta_B [\bar C^a(\frac {1}{2} h_{ab})] = -i \delta_B [\bar C^a(\frac {1}{2}A_a)]
\label{cfgf}
\end{eqnarray}
where $\lambda$ is infinitesimal, anti-commuting, global BRST parameter.
Total Lagrangian density in conformal gauge in extended form is then written as
\begin{eqnarray}
\mathscr{L}_{cf}^t &=& \frac{1}{2}\sqrt{-g} g^{ab}\partial_a X^\mu \partial_b X_\mu - \frac{1}{2}\sqrt{-\hat g} b^a A_a - \frac{1}{2}\sqrt{-\hat g} \delta_B (A_a)\bar C^a
\label{taccf}
\end{eqnarray}
where $b^a$ is Nakanishi-Lautroop type auxiliary field. 

\subsection{Harmonic Gauge}\label{Harmonic Gauge}

The harmonic gauge condition for bosonic string action is written as\cite{7,8,9}
\begin{equation}
\hat{\nabla}_a (\sqrt{-g} g^{ab})  = 0
\label{hgc}
\end{equation} 
where the Christoffel connection in $\hat{\nabla}$ is calculated for an arbitrary background metric $g^{ab}$.

In order to calculate gauge-fixing and ghost part of the action we split the metric $g_{ab}$ into a classical background field $\hat{g}_{ab}$ and a (quantum) metric perturbation $h_{ab}$ as
\begin{eqnarray}
g_{ab} = \hat{g}_{ab} + h_{ab}
\label{mpt}
\end{eqnarray}

We need to fix the gauge condition only for the quantum field $h_{ab}$  
\begin{eqnarray}
\delta h_{ab} &=& \hat{\nabla}_a \xi_b + \hat{\nabla}_b \xi_a + h_{bc}\hat{\nabla}_a \xi^c + h_{ac}\hat{\nabla}_b \xi^c + \xi^c \hat{\nabla}_c h_{ab}\nonumber\\
\delta h_{ab} &=& 2\sigma (\hat{g}_{ab} + h_{ab})  
\label{gf}
\end{eqnarray}
here the classical field $\hat{g}_{ab}$ is invariant under general coordinate transformation together with standard tensor transformation rules for the other fields.

The linearized form of gauge fixing condition in harmonic gauge is
\begin{eqnarray}
\frac {1}{2}\hat{\nabla}_a h - \hat{\nabla}^b h_{ab} = 0
\label{lhgc} 
\end{eqnarray}
 where $h = \hat{g}^{ab} h_{ab}$.
The gauge condition in Eq.(\ref{lhgc}) as well as in Eq.(\ref{hgc}) are not quantum Weyl covariant. To make them Weyl covariant we need a special gauge fixing condition.
The gauge fixing condition is given by 
\begin{eqnarray}
\hat{g}^{ab} h_{ab} = 0
\label{sgfc} 
\end{eqnarray}

Now the gauge fixing and ghost part of the action can be written in BRST invariant manner as,
\begin{eqnarray}
\mathscr{L}_{gf} + \mathscr{L}_{gh} = \delta_B[\bar C^a(\frac {1}{2}\hat{\nabla}_a h - \hat{\nabla}^b h_{ab}) + \bar \tau \hat{g}^{ab} h_{ab}]
\label{gffp} 
\end{eqnarray}
where $\bar \tau$ is Weyl antighost field.

This action can be further simplified using the technique in Ref.{\cite{8}}. The simplified gauge fixing and ghost term can be written in BRST invariant manner as,
\begin{eqnarray}
\mathscr{L}_{hm} \equiv \lambda(\mathscr{L}_{gf} + \mathscr{L}_{gh}) = -i\delta_B[\bar C^a(\frac {1}{2}\hat{\nabla}_a h - \hat{\nabla}^b h_{ab})]
\label{hmgf} 
\end{eqnarray} 
Total Lagrangian density in background covariant harmonic gauge is then written in extended form as
\begin{eqnarray}
\mathscr{L}_{hm}^t &=& \frac{1}{2}\sqrt{-g} g^{ab}\partial_a X^\mu \partial_b X_\mu - \sqrt{-\hat g} b^a \hat{\nabla}^b h_{ab} - i \sqrt{-\hat g}[\hat \nabla^b {\bar C}^a \hat \nabla_b C_a - {\bar C}^a \hat R_{ab}C^b  + (\hat \nabla^b {\bar C}^a + \hat \nabla^a {\bar C}^b\nonumber \\ &-& {\hat g}^{ab}\nabla . {\bar C})h_{bc}\hat \nabla_a C^c + \hat \nabla^b {\bar C}^a C^c \hat \nabla_c h_{ab} - h_{ab} \hat \nabla^a {\bar C}^b (\hat\nabla . {\bar C} + h_{ij}\hat \nabla^i  C^j) ] 
\label{tachm} 
\end{eqnarray} 
where $b^a$ is Nakanishi-Lautroop type auxiliary field.
 
\subsection{BRST Symmetry}\label{BRST Symmetry}
The total action is defined as 
\begin{eqnarray}
S_{0}^t = \int d^2 x(\mathscr {L}_0 + \mathscr {L}_{gf} + \mathscr {L}_{gh})
\label{st}
\end{eqnarray}
where $\mathscr {L}_0$ is the kinetic part of the total Lagrangian density. This total action in Eqs.(\ref{taccf},\ref{tachm}) are invariant under following BRST transformation\cite{8},
\begin{eqnarray}
\delta_B h_{ab} &=& i\lambda[\hat{\nabla}_a C_b + \hat{\nabla}_b C_a + h_{bc}\hat{\nabla}_a C^c + h_{ac}\hat{\nabla}_b C^c + C^c \hat{\nabla}_c h_{ab}- (\hat \nabla .C + h_{ij}\hat \nabla^i C^j)(h_{ab} + \hat g_{ab})],\nonumber\\
\delta_B X^\mu &=& i\lambda C^a \partial_a X^\mu,\quad \delta_B C^a = i\lambda C^b \hat{\nabla}_b C^a, \quad \delta_B \bar C^a = \lambda b^a,\quad  \delta_B b^a = 0
\label{brst}
\end{eqnarray}
where $\lambda$ is infinitesimal, anti-commuting, global BRST parameter. One can verify that these transformations are nilpotent.

\section{Connection between generating functionals in Background Covariant Harmonic and Conformal gauges}\label{Connection between generating functionals in Background Covariant Harmonic and conformal gauges}

Before going to show the connection between the two effective theories we briefly discuss the ideas of FFBRST developed in Ref.\cite{13}. The BRST transformations  are generated from BRST charge using relation $\delta_b \phi = -[Q,\phi]\lambda$ where $\lambda$ is infinitesimal anti-commuting global parameter. Following the technique in  Ref.\cite{13} the anti-commuting BRST parameter $\lambda$ is generalized to be finite-field dependent but space time independent parameter $\Theta[\phi]$. Since the parameter is finite in nature unlike the usual case the path integral measure is not invariant under such finite transformation.  The Jacobian for these transformations for certain $\Theta[\phi]$ can be calculated by summing the Jacobian contribution of infinitesimal but field dependent BRST transformation $\lambda = \Theta'[\phi(k)]dk$  
\begin{eqnarray}
D\phi &=& J(k)D\phi'(k)\nonumber\\
       &=& J(k+dk)D\phi'(k+dk)
\label{chj}
\end{eqnarray}
where  a  numerical parameter $k$ ($0\leq k \leq 1$), has been introduced  to execute the finite transformation in a mathematically convenient way. All the fields are taken to be $k$ dependent in such a fashion that $\phi(x,0) = \phi(x)$ and $\phi(x,k = 1) = \phi'(x)$. 
$S_{eff}$ is invariant under FFBRST which is constructed by considering successive infinitesimal BRST transformations $(\phi(k)\rightarrow\phi(k+dk))$. The nontrivial Jacobian $J(k)$ is then written as local functional of fields and will be replaced as $e^{iS_1[\phi(k),k]}$ if the condition 

\begin{eqnarray}
\int D\phi(k)\big[\frac{1}{J(k)}\frac{d J(k)}{d k}-i\frac{dS_1}{dk}\big] e^{i (S_1 + S_{eff})} = 0 
\label{ffbc}
\end{eqnarray}
holds \cite{13}. Where $\frac{dS_1}{dk}$ is a total derivative of $S_1$ with respect to $k$ in which dependence on $\phi(k)$ is also differentiated. The change in Jacobian is calculated as 
\begin{eqnarray}
\frac{J(k)}{j(k+dk)} &=& \Sigma_{\phi}{\pm}\frac{\delta \phi(x,k+dk)}{\delta \phi(x,k)}\nonumber\\
                     &=& 1 - \frac{1}{J(k)}\frac{d J(k)}{d k} d k 
\label{echj}
\end{eqnarray}
${\pm}$ is for bosonic and fermionic fields respectively.The Eq.(\ref{ffbc}) further can be simplified as 
\begin{eqnarray}
\frac{1}{J(k)}\frac{dJ(k)}{dk} = \Sigma_{\phi}{\pm}\delta_B \phi \frac{\partial\Theta'}{[\partial \phi ]}
\label{jacth}
\end{eqnarray}
where $\Theta'[\phi]$ is used to construct the finite parameter $\Theta = \int_{0}^k \Theta' dk $.
In this section, we construct the FFBRST transformation with an appropriate finite parameter to obtain the generating functional corresponding to $\mathcal{L}_{hm}^t$ from that of  corresponding to $\mathcal{L}_{cf}^t$. We calculate the Jacobian corresponding to such a FFBRST transformation following the method outlined in Ref.\cite{13} and show that it is a local functional of fields. We construct out FFBRST parameter in such a way that Jacobian contribution accounts for the differences of the two FP effective actions.

 Now we start with the generating functional corresponding to the FP effective action in conformal gauge is written as
\begin{eqnarray}
Z_{cf}^t = \int D\phi \exp (iS_{cf}^t[\phi])
\label{gfcfg}
\end{eqnarray}
where $S_{cf}^t$ is given by
\begin{eqnarray}
S_{cf}^t = \int d^2 x [\mathscr {L}_0 + \mathscr{L}_{cf}]
\label{shm}
\end{eqnarray}
Now, to obtain the generating functional corresponding $S_{hm}^t$, we  apply the FFBRST transformation with a finite parameter $\Theta[\phi]$ which is obtained from the infinitesimal but field dependent parameter, $\Theta'[\phi(k)]$ through $ \int_0^\kappa \Theta^\prime[\phi(\kappa)] d\kappa$, we construct $ \Theta^\prime[\phi(\kappa)] $ as,
\begin{eqnarray}
\Theta'[C, h] = i \int d^2 x [\gamma \bar C^a \{\frac {1}{2} A_a -(\frac {1}{2}\hat{\nabla}_a h - \hat{\nabla}^b h_{ab}) \} ]
\label{thd}
\end{eqnarray}
Here $\gamma$ is arbitrary constant parameter and all the fields depend on the parameter $k$.
The infinitesimal change in the Jacobian corresponding to this FFBRST transformation is calculated using Eq.(\ref{jacth}) 
\begin{eqnarray}
\frac{1}{J(k)}\frac{d J(k)}{d k} = -i \int d^2 x \gamma[\delta(\bar C^a)\{ \frac {1}{2} A_a -( \frac {1}{2}\hat{\nabla}_a h - \hat{\nabla}^b h_{ab})\} + \frac {1}{2}\delta (A_a)\bar C^a - \frac {1}{2}\hat{\nabla}^a\delta h \bar C^a + \hat{\nabla}^b \delta (h_{ab})\bar C^a] 
 \label{chij}
\end{eqnarray}
To express the Jacobian contribution in terms of a local functional of fields, we make an ansatz for $S_1$ by considering all possible terms that could arise from such a transformation as
\begin{eqnarray}
S_1[\phi(k), k]  &=&  \int d^2 x [ \frac {\xi_1}{2}\delta(\bar C^a) A_a + \frac{\xi_2}{2} \delta(\bar C^a)\hat{\nabla}_a h  + \xi_3 \delta(\bar C^a)\hat{\nabla}^b h_{ab}  + \frac {\xi_4}{2}\delta (A_a)\bar C^a \nonumber\\ &+& \frac {\xi_5}{2}\hat{\nabla}_a \delta h(\bar C^a) + \xi_6\hat{\nabla}^b \delta (h_{ab})\bar C^a ]  
\label{son}
\end{eqnarray}
where all the fields are considered to be $k$ dependent and we have introduced arbitrary $k$ dependent parameters $\xi_n=\xi_n(k) (n =1, 2, .., 6)$ with initial condition $\xi_n(k = 0) = 0$. It is straight forward to calculate
\begin{eqnarray}
\frac{dS_1}{dk} &=& \int d^2 x [\frac {\xi'_1}{2}\delta(\bar C^a) A_a + \frac{\xi_2'}{2}\delta(\bar C^a)\hat{\nabla}_a h  + \xi'_3 \delta(\bar C^a)\hat{\nabla}^b h_{ab}  + \xi'_4\frac {1}{2}\delta (A_a)\bar C^a + \frac {\xi'_5}{2}\hat{\nabla}_a \delta h(\bar C^a)\nonumber\\ &-& \xi'_6 \hat{\nabla}^b \delta (h_{ab})\bar C^a + \Theta'\{ \frac {\xi_1}{2}\delta (A_a)\delta(\bar C^a) + \frac{\xi_2}{2}\hat{\nabla}_a \delta h\delta(\bar C^a) + \xi_3 \hat{\nabla}^b \delta h_{ab}\delta(\bar C^a) + \frac {\xi_4}{2}\delta(\bar C^a)\delta (A_a) \nonumber\\ & + & \frac {\xi_5}{2}\delta(\bar C^a)\hat{\nabla}_a\delta h + \xi_6\delta(\bar C^a)\hat{\nabla}^b \delta (h_{ab}) \} ] 
\label{dsdk}
\end{eqnarray} 

where $\xi'_n \equiv \frac {d\xi_n}{dk}$.
Now to satisfy the condition in Eq.(\ref{ffbc}).
\begin{eqnarray}
 &\int  D\phi &  \exp[{i (S_{CF}[\phi(k) ] + S_1[\phi(k), k] )}]\int d^2 x [\frac {(\gamma + \xi'_1 )}{2}\delta(\bar C^a) A_a + \frac{(-\gamma + \xi'_2 )}{2}\delta(\bar C^a)\hat{\nabla}_a h\nonumber\\ &+& (\gamma + \xi'_3 )\delta(\bar C^a)\hat{\nabla}^b h_{ab} + \frac{(\gamma + \xi'_4 )}{2}\delta (A_a)\bar C^a  + \frac{(-\gamma + \xi'_5 )}{2}\hat{\nabla}_a \delta h(\bar C^a) + (\gamma + \xi'_6)\hat{\nabla}^b \delta (h_{ab})\bar C^a  \nonumber\\ &+& \Theta'\{ \frac{(\xi_1 - \xi_4)}{2}\delta (A_a)\delta(\bar C^a) + \frac{(\xi_2 - \xi_5 )}{2} \hat{\nabla}_a \delta h\delta(\bar C^a) + (\xi_3 -  \xi_6 )\hat{\nabla}^b \delta h_{ab}\delta(\bar C^a) \} ] = 0
 \label{ffbrc}
\end{eqnarray} 

The terms proportional to $\Theta'$, which are nonlocal due to $\Theta'$, vanish independently  if

\begin{eqnarray}
\xi_1 - \xi_4 = 0, \quad \xi_2 - \xi_5 = 0, \quad \xi_3 - \xi_6 = 0
\label{reiji}
\end{eqnarray}
To make the remaining local terms in Eq.(\ref{ffbrc}) vanish, we need the following conditions:

\begin{eqnarray}
\gamma + \xi'_1 = 0, \quad -\gamma + \xi'_2 = 0, \quad \gamma + \xi'_3 &=&0 \nonumber\\
\gamma + \xi'_4 = 0, \quad -\gamma + \xi'_5 = 0, \quad \gamma + \xi'_6 &=&0
\label{rbjg}
\end{eqnarray}
The differential equations for $\xi_n(k)$ can be solved with the initial conditions $\xi_n(0) = 0$ to obtain the solutions
\begin{eqnarray}
\xi_1 = -\gamma k, \quad  \xi_2 = \gamma k,\quad \xi_3 = -\gamma k,\quad \xi_4 = -\gamma k,\quad \xi_5 = \gamma k, \quad \xi_6 = -\gamma k 
\label{voj}
\end{eqnarray}

Putting values of these parameters in expression of $S_1$, and choosing arbitrary parameter $\gamma = 1$ we obtain,
\begin{eqnarray}
S_1[\phi(1), 1] &=& \int d^2 x [- \frac {1}{2}\delta(\bar C^a) A_a + \frac{1}{2} \delta(\bar C^a)\hat{\nabla}_a h - \delta(\bar C^a)\hat{\nabla}^b h_{ab}  
- \frac {1}{2}\delta (A_a)\bar C^a + \frac {1}{2}\hat{\nabla}_a \delta h(\bar C^a)\nonumber\\ &-& \hat{\nabla}^b \delta (h_{ab})\bar C^a ] \nonumber\\
&=& \int d^2 x [\frac {1}{2}b^a A_a - \frac{1}{2} b^a \hat{\nabla}_a h + b^a \hat{\nabla}^b h_{ab} - \frac {1}{2}\delta (A_a)\bar C^a + \frac {1}{2}\hat{\nabla}_a g^{bc} \{\hat{\nabla}_b C_c + \hat{\nabla}_c C_b + h_{cd}\hat{\nabla}_b C^d \nonumber\\ &+& h_{bd}\hat{\nabla}_c C^d + C^d \hat{\nabla}_d h_{bc} - (\hat \nabla .C + h_{ij}\hat \nabla^i C^j)(h_{bc} + \hat g_{bc})\} {\bar C}^a - \hat{\nabla}^b \{\hat{\nabla}_a C_b + \hat{\nabla}_b C_a \nonumber\\ &+& h_{bc}\hat{\nabla}_a C^c + h_{ac}\hat{\nabla}_b C^c + C^c \hat{\nabla}_c h_{ab} - (\hat \nabla .C + h_{ij}\hat \nabla^i C^j)(h_{ab} + \hat g_{ab})\}\bar C^a ]
\label{yeso}
\end{eqnarray}
Thus the FFBRST transformation with the finite parameter $\Theta$ that is defined by Eq.(\ref{thd}) changes the generating functional $Z_{CF}$ as
\begin{eqnarray}
Z_{cf}^t &=& \int D\phi \exp (iS_{cf}^t[\phi ] )\nonumber\\
&=&\int D\phi'\exp[{i (S_{cf}^t[\phi' ] + S_1[\phi', 1] )}] \nonumber\\
&=&\int D\phi\exp[{i (S_{cf}^t[\phi ] + S_1[\phi , 1] )}] 
\end{eqnarray}
which can be written as
\begin{eqnarray}
Z_{cf}^t&=&\int D\phi\exp [i\int d^2 x \{\frac{1}{2}\sqrt{-g} g^{ab}\partial_a X^\mu \partial_b X_\mu - \frac{1}{2}\sqrt{-\hat g} b^a A_a - \frac{1}{2}\sqrt{-\hat g} \delta_B (A_a)\bar C^a + \frac {1}{2}b^a h_{ab} - \frac{1}{2} b^a \hat{\nabla}_a h \nonumber\\ &+& b^a \hat{\nabla}^b h_{ab} - \frac {1}{2}\delta (A_a) \bar C^a + \frac {1}{2}\hat{\nabla}_a g^{bc} \{\hat{\nabla}_b C_c + \hat{\nabla}_c C_b + h_{cd}\hat{\nabla}_b C^d + h_{bd}\hat{\nabla}_c C^d + C^d \hat{\nabla}_d h_{bc}\nonumber\\ &-& (\hat \nabla .C + h_{ij}\hat \nabla^i C^j)(h_{bc} + \hat g_{bc})\}{\bar C}^a - \hat{\nabla}^b \{\hat{\nabla}_a C_b + \hat{\nabla}_b C_a + h_{bc}\hat{\nabla}_a C^c + h_{ac}\hat{\nabla}_b C^c + C^c \hat{\nabla}_c h_{ab} \nonumber\\ &-& (\hat \nabla .C + h_{ij}\hat \nabla^i C^j)(h_{ab} + \hat g_{ab})\}\bar C^a\} ]
\label{cfson}
\end{eqnarray}
After sum simplification we will get
\begin{eqnarray}
Z_{cf}^t &=& \int D\phi [i\int d^2 x \{\exp\frac{1}{2}\sqrt{-g} g^{ab}\partial_a X^\mu \partial_b X_\mu - \sqrt{-\hat g} b^a \hat{\nabla}^b h_{ab} - i \sqrt{-\hat g}[\hat \nabla^b {\bar C}^a \hat \nabla_b {\bar C}_a - {\bar C}^a \hat R_{ab}C^b  + (\hat \nabla^b {\bar C}^a \nonumber \\ & + &\hat \nabla^a {\bar C}^b - {\hat g}^{ab}\nabla . {\bar C})h_{bc}\hat \nabla_a C^c + \hat \nabla^b {\bar C}^a C^c \hat \nabla_c h_{ab} - h_{ab} \hat \nabla^a {\bar C}^b (\hat\nabla . {\bar C} + h_{ij}\hat \nabla^i  C^j)\} ] 
\end{eqnarray}
or
\begin{eqnarray}
Z_{cf}^t = \int D\phi \exp (iS_{hm}^t[\phi ] ) \equiv Z_{hm}^t
\label{acf}
\end{eqnarray}
Here $S_{hm}^t$ is defined as
\begin{eqnarray}
S_{hm}^t = \int d^2 x(\mathcal {L}_x + \mathcal {L}_{hm})
\label{scf}
\end{eqnarray}
In this way FFBRST transformation with the finite field dependent parameter in Eq.(\ref{thd}) connects generating functional for the Polyakov action in the conformal gauge to that of in background covariant harmonic gauge, where ghost number anomaly does not appear. Thus we can avoid the ghost number anomaly by using suitably constructed FFBRST transformation. 
\section{Conclusion}\label{Conclusion}
In this present work we have shown how ghost number current anomaly present in conformal gauge in curved worldsheet is removed using field transformation. By constructing appropriate FFBRST transformation we obtain the generating functional in conformal gauge from that of in harmonic gauge. This provides a convenient way to go from a theory with ghost number anomaly to the theory where there is no ghost number current anomaly. Further harmonic gauge which is complicated is directly connected through the constructed field transformation to conformal gauge theory which is simpler to use. It will be interesting to generalize this formulation for superstring theories.

One of us (VKP) acknowledges the University Grant Commission (UGC), India, for its financial assistance under the CSIR-UGC JRF/SRF scheme. BPM ask support from CAS program of Department of physics, BHU.

\end{document}